\documentclass[10pt, journal]{IEEEtran}

\usepackage{authblk}
\usepackage{graphicx}
\usepackage{subcaption}
\usepackage{xcolor}
\usepackage{acronym}
\usepackage{amsmath}
\usepackage{amssymb}
\usepackage{cite}
\usepackage{bm}
\usepackage{fancyhdr}
\makeatletter
\def\@citex[#1]#2{\leavevmode
\let\@citea\@empty
\@cite{\@for\@citeb:=#2\do
{\@citea\def\@citea{,\penalty\@m\ }%
\edef\@citeb{\expandafter\@firstofone\@citeb\@empty}%
\if@filesw\immediate\write\@auxout{\string\citation{\@citeb}}\fi
\@ifundefined{b@\@citeb}{\hbox{\reset@font\bfseries ?}%
\ G@refundefinedtrue
\@latex@warning
{Citation `\@citeb' on page \thepage \space undefined}}%
{\@cite@ofmt{\csname b@\@citeb\endcsname}}}}{#1}}
\makeatother
\usepackage{soul,color}

\acrodef{UE}[UE]{user equipment}
\acrodef{GNN}[GNN]{graph neural network}
\acrodef{RL}[RL]{Reinforcement Learning}
\acrodef{MADRL}[MADRL]{Multi-agent deep reinforcement learning}
\acrodef{MARL}[MARL]{multi-agent reinforcement learning}
\acrodef{RRM}[RRM]{Radio Resource Management}
\acrodef{WLAN}[WLAN]{wireless local area network}
\acrodef{CTDE}[CTDE]{centralized training decentralized execution}
\acrodef{AP}[AP]{Access Point}
\acrodef{MLP}[MLP]{Multi-Layer Perceptron}
\acrodef{k-GNN}[local $k$-GNN]{Local $k$-dimensional GNN}
\acrodef{MNO}[MNO]{Mobile Network Operator}
\acrodef{gNB}[gNB]{gNodeB}
\acrodef{CCO}[CCO]{Capacity and Coverage Optimization}
\acrodef{PPP}[PPP]{homogeneous Poisson point process}
\acrodef{RSRP}[RSRP]{Reference Signal Received Power}
\acrodef{SINR}[SINR]{signal to interference plus noise ratio}
\acrodef{DDPG}[DDPG]{Deep Deterministic Policy Gradient}
\acrodef{POMDP}[POMDP]{partially observable Markov decision process}
\acrodef{PPO}[PPO]{Proximal Policy Optimization}
\acrodef{STAR-RISs}[STAR-RISs]{Simultaneously Transmitting and Reflecting Reconfigurable Intelligent Surfaces}
\acrodef{HetNet}[HetNet]{Heterogeneous Network}
\acrodef{BER}[BER]{bit error rate}
\acrodef{PCP}[PCP]{Poisson cluster process}
\acrodef{MLB}[MLB]{mobility load balancing}
\title{Multi-Agent Reinforcement Learning for Power Control in Wireless Networks via Adaptive Graphs}

\author[1, 3, *]{Lorenzo Mario Amorosa}
\author[1, 3, *]{Marco Skocaj}
\author[1, 3]{Roberto Verdone}
\author[2, 3]{Deniz Gündüz}
\affil[1]{Department of Electrical, Electronic and Information Engineering, University of Bologna, \textit{Italy}}
\affil[2]{Department of Electrical and Electronic Engineering, Imperial College London, \textit{U.K.}}
\affil[3]{WiLab - National Wireless Communication Laboratory (CNIT), \textit{Italy}}
\affil[*]{\textit {Corresponding authors: \{lorenzomario.amorosa, marco.skocaj\}@unibo.it}}

\begin{document}
\bstctlcite{IEEEexample:BSTcontrol}

\maketitle
% remove page numbers
\thispagestyle{empty}

\begin{abstract}
The ever-increasing demand for high-quality and heterogeneous wireless communication services has driven extensive research on dynamic optimization strategies in wireless networks. Among several possible approaches, multi-agent deep reinforcement learning (MADRL) has emerged as a promising method to address a wide range of complex optimization problems like power control. However, the seamless application of MADRL to a variety of network optimization problems faces several challenges related to convergence. In this paper, we present the use of graphs as communication-inducing structures among distributed agents as an effective means to mitigate these challenges. Specifically, we harness graph neural networks (GNNs) as neural architectures for policy parameterization to introduce a relational inductive bias in the collective decision-making process. Most importantly, we focus on modeling the dynamic interactions among sets of neighboring agents through the introduction of innovative methods for defining a graph-induced framework for integrated communication and learning. Finally, the superior generalization capabilities of the proposed methodology to larger networks and to networks with different user categories is verified through simulations.
\end{abstract}

\begin{IEEEkeywords}
Graph Neural Networks, Multi-Agent Deep Reinforcement Learning, Wireless Networks, Power Control
\end{IEEEkeywords}

\section{Introduction}
\label{sec:intro}

Wireless communication networks constitute complex systems demanding careful optimization of network procedures to attain predefined performance objectives. \ac{MADRL}, owing to its inherent advantages, has emerged as a promising strategy for the optimization of a variety of network problems. Nevertheless, the practical implementation of \ac{MADRL} in real systems is hindered by challenges related to convergence, which continue to constitute an active area of research. These challenges encompass the non-stationarity of the environment, the partial observability of the state, as well as the coordination and cooperation among agents \cite{zhang2021multi, gronauer2022multi}. To this end, this paper elucidates the role of leveraging graph structures as an effective means to account for non-stationarity in \ac{MADRL} systems by introducing a relational inductive bias in the collective decision-making process. Leveraging \acp{GNN} as neural architectures for policy optimization, the learning process is governed by feature aggregation over neighbor entities, and computations over graphs afford a strong relational inductive bias beyond what convolutional and recurrent layers can provide \cite{battaglia2018relational}. The underscoring intuition behind this approach is that choosing the architecture with the right bias for the problem at hand might dramatically increase cooperation among agents, allowing them to communicate local features to the relevant peers to compensate for partial observability. Moreover, \ac{GNN}-based optimization is capturing significant interest for cellular networks, as peer-to-peer communication between base stations and AI-related messaging over the control plane are currently being studied for standardization \cite{lin2023artificial}.

In this work, graph structures are harnessed to tackle a power control optimization problem in cellular networks, serving as an illustrative example within the realm of network optimization in mobile radio networks. A particular emphasis is placed on effectively accounting for the mutual interplay between sets of neighboring agents through the introduction of innovative strategies for defining a graph-induced framework for integrated communication and learning.

\begin{figure*}[!ht]
    \centering
    \includegraphics[width = 0.8\textwidth]{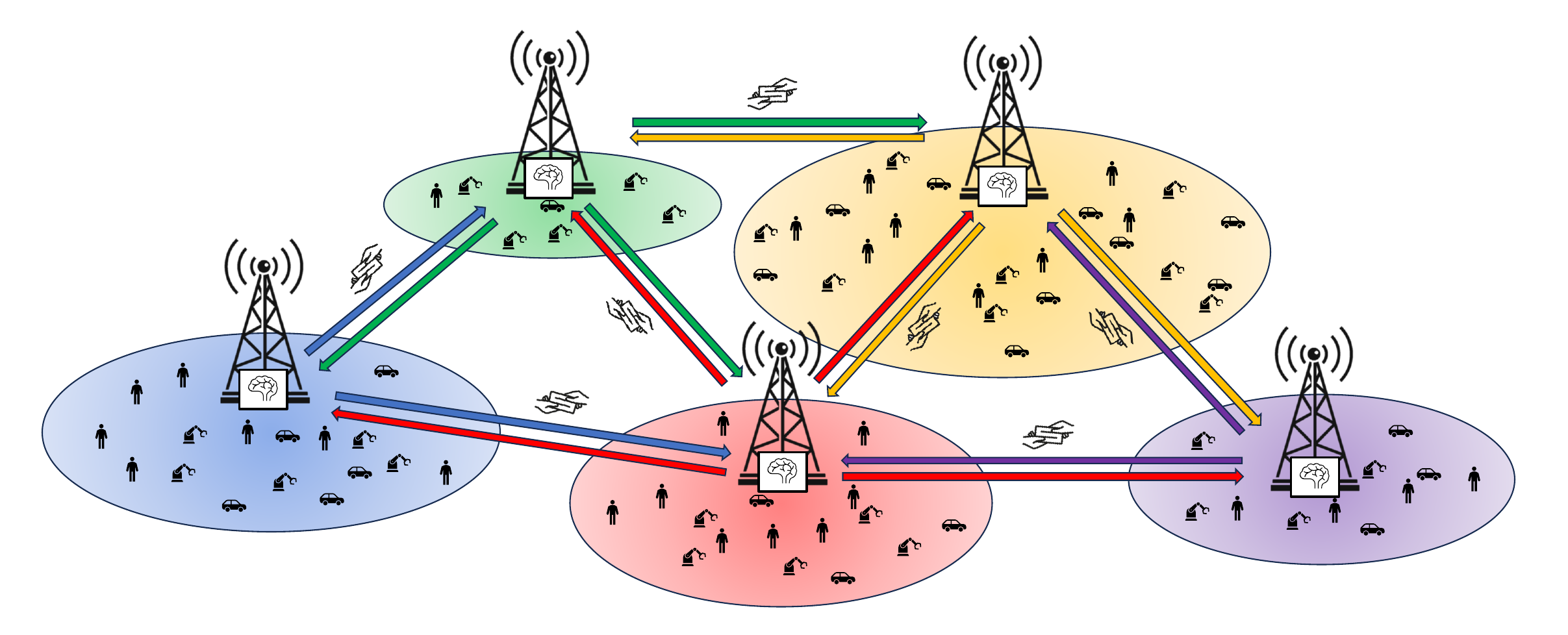}
    \caption{Base stations perform power control optimization in a decentralized manner by relying on a communication graph that enables the exchange of information exclusively among connected nodes.}
    \label{fig:reference_scenario}
    \vspace{-1em}
\end{figure*}

%%%% START ARXIV
\thispagestyle{fancy}
\fancyhf{}
\fancyfoot[C]{\textit{This work has been submitted to the IEEE for possible publication. Copyright may be transferred without notice, after which this version may no longer be accessible.}}
\renewcommand{\headrulewidth}{0pt}
%%%% END ARXIV

\section{Related Work and Contributions}
\label{sec:literature}

The utilization of \acp{GNN} for the optimization of wireless networks has gained significant traction in the recent literature. This interest can be attributed to the innate characteristics of GNNs, which enable a scalable solution and exhibit inductive capability and, thanks to the permutation equivariance property, increased generalization. Notably, these properties find practical application in works such as \cite{eisen2020optimal}, where GNNs are harnessed to capture the dynamic structure of fading channel states for the purpose of learning optimal resource allocation policies in wireless networks. Another domain that has witnessed substantial utilization of \acp{GNN} is channel management within wireless local area networks (WLANs), as evidenced by works such as \cite{gao2022decentralized} and \cite{nakashima2020deep}. A notable insight derived from the study by Gao et al. \cite{gao2022decentralized} is the inherent property of GNNs to provide decentralized inference, rendering them a viable and promising approach for the practical implementation of over-the-air \ac{MADRL} systems. Similarly, the application of GNNs in addressing power control optimization challenges within wireless networks is explored in works such as \cite{Ribeiro2022Adaptive}, \cite{zhang2021scalable}, and \cite{li2023heterogeneous}.

Throughout the body of pertinent literature, GNNs are utilized as either centralized controllers or decentralized entities to model data-driven policies based on feature convolution over graphs. However, despite the crucial role of graph structure in defining agents' interactions, the effect of different graph formation strategies on achieving a collective goal is a problem often overlooked. Instead, the definition of the graph structure has an influence on inducing communication among distributed agents, thereby enabling effective cooperation. For instance, \cite{jiang2018learning} puts forth the idea that uncontrolled information sharing among all agents in a distributed setting could be detrimental to the learning process. To this end, an attentional communication model is proposed to learn \textit{when} communication should take place. Another noteworthy work in this field is presented by \cite{das2019tarmac}, showcasing how targeted communication, where agents learn both \textit{what} messages to send and \textit{whom} to address them, can compensate for partial observability in a multi-agent setting.

In light of the above background, the presented work introduces the following contributions:

\begin{itemize}
    \item We consider graphs as communication-inducing structures for distributed optimization problems. Specifically, by parameterizing a joint action policy leveraging \acp{GNN}, the relational inductive bias introduced by message transformation (i.e., \textit{what} to communicate) and message passing (i.e., \textit{whom} to communicate) is shown to have a deep impact on the decision-making process.
    \item We showcase increasingly articulated strategies to integrate domain knowledge into the formulation of graph construction strategies. Our investigation demonstrates the substantial impact of these strategies on the collective capacity to acquire cooperative behavior and enhance the overall performance.
    \item Finally, a novel approach to learning optimal edge weights in an end-to-end fashion is presented, showcasing impressive inductive capabilities and surpassing conventional strategies grounded in domain expertise.
\end{itemize}

% remove page numbers
\thispagestyle{empty}

\section{System Model and Formulation}
\label{sec:sys_mod}

The reference scenario under consideration includes a group of base stations concurrently providing communication services to a set of \acp{UE}, categorized based on their distinct service and performance requirements. Specifically, $S$ categories are considered, with increasingly demanding requirements in terms of reliability (\acf{BER}). An overall depiction of the scenario is provided in Fig. \ref{fig:reference_scenario}. Wireless network's modeling, \acp{UE}' requirements and traffic distributions, and the \ac{POMDP} formulation are described hereafter.

\subsection{Wireless network modeling}
\label{sec:wnet}
We consider a graph representation of the network $\mathcal{G}=(\mathcal{V}, \mathcal{E})$, where nodes $\mathcal{V}$ correspond to base stations/agents\footnote{We use the terms base stations and agents interchangeably.} and edges $\mathcal{E}$ represent virtual links between them. To determine the graph structure, it is necessary to define an adjacency matrix, denoted by $A \in \mathbb{R}^{|\mathcal{V}|\times|\mathcal{V}|}$. Each element $a_{u,v} \in A$ indicates the connectivity between nodes $u, v \in \mathcal{V}$.

A fundamental aspect of the analysis presented here revolves around defining the proper graph-inducing structure for the set of base stations. To this end, four distinct strategies to determine $A$ are presented.

\subsubsection{Binary edges}
As a first strategy, we consider a binary representation for the edges:

\begin{equation}
\label{eq:binary_weights}
    a_{u,v} = \begin{cases}
            1 & \text{if $||\mathbf{s}(u) - \mathbf{s}(v)||_2 < D$ } \\
            0 & \text{otherwise}
        \end{cases}~,
\end{equation}
where $\mathbf{s}(u)$ denotes the position of node $u$ in the 2D-space, $|| \cdot ||_2$ is the Euclidean distance, and $D \in \mathbb{R}$ is a threshold. This results in a symmetric matrix, and the corresponding graph $\mathcal{G}(\mathcal{V}, \mathcal{E})$ is \textit{unweighted} and \textit{undirected}.\\

\subsubsection{Distance-based edges}
A more informative approach involves considering edges with continuous values $a_{u,v} \in \mathbb{R}$, based on the physical proximity between the nodes, i.e., 
\begin{equation}
\label{eq:distance_based_weights}
    a_{u,v} \propto e^{-||\mathbf{s}(u) - \mathbf{s}(v)||_2}~.
\end{equation}
This results in a \textit{weighted} and \textit{undirected} graph, since the physical distance between two nodes is symmetric.\\

\subsubsection{Relation-based edges} This method involves determining edges based on the mutual interaction between the sets of nodes. In other words, nodes are deemed adjacent if their actions have a mutual influence on one another. The evaluation of this mutual interaction poses a non-trivial challenge and necessitates the application of advanced tools such as directed graphical modeling \cite{Jordan2004Graphical, GoodBengCour16}, causal inference \cite{Pearl2010Causal}, or, in the specific context of power optimization in wireless networks, the utilization of measurements or empirical models to quantify and assess mutual interference. In a general formulation, this can be expressed as

\begin{equation}
\label{eq:relation_based_weights}
    a_{u,v} \propto \mathcal{I}(u \vert v)~,
\end{equation}
where $\mathcal{I}(u \vert v)$ indicates the influence that node $v$ exerts on $u$, and typically $\mathcal{I}(u \vert v) \not = \mathcal{I}(v \vert u)$. In a wireless communication network, $\mathcal{I}(u \vert \cdot)$ could be determined as the level of interfering power at node $u$ from each node $v \in \mathcal{V}$ connected to $u$.
This results in a \textit{weighted} and \textit{directed} or \textit{undirected} graph, depending on the symmetric nature of $\mathcal{I(\cdot \vert \cdot)}$.\\

\subsubsection{Learning-based edges}
In our final strategy, an end-to-end learning method is introduced for the determination of $A$. During the collective training phase of agents, the graph's edge weights are derived simultaneously with the policy parameters. This is achieved by using network topological characteristics as input features for a separate \ac{GNN}. This second \textit{auxiliary} GNN is not tasked with parameterizing the policy, but focuses on the dynamic assignment of edge weights as the training progresses.

The initial stage of the process entails determining the geometric properties and connections among network nodes. To accomplish this, we establish edge features, denoted as $\mathcal{F}_{u,v}$, for each directed edge $a_{u,v}$. The edge feature is denoted by a set composed as follows:

\begin{equation}
\label{eq:edge_features}
    \mathcal{F}_{u,v} = \{d_{u,v},~\sin(\theta_{u,v}),~\cos(\theta_{u,v})\}~,
\end{equation}
where:
\begin{itemize}
    \item $d_{u,v} \triangleq \| u - v \|_2$ is the Euclidean distance between nodes $u$ and $v$.
    \item $\theta_{u,v} \triangleq \arctan2(v_y - u_y, v_x - v_x)$ is the angle between nodes $u$ and $v$.
\end{itemize}
The training of the auxiliary GNN for edge prediction hinges upon the introduction of an auxiliary graph structure, denoted as $\mathcal{G}_f=(\mathcal{V}_f, \mathcal{E}_f)$. In this representation, nodes $\mathcal{V}_f$ correspond to the set of edges $\mathcal{E}$ of the original graph $\mathcal{G}$, and node features on $\mathcal{G}_f$ are the corresponding edge features $\mathcal{F}_{u,v}$ of $\mathcal{G}$. The set of edges $\mathcal{E}_f$ connecting nodes in $\mathcal{G}_f$ is determined as a binary adjacency matrix. Since nodes $\mathcal{V}_f$ in $\mathcal{G}_f$ correspond to edges $\mathcal{E}$ in $\mathcal{G}$, two nodes in $\mathcal{V}_f$ are deemed adjacent if their associated edges in $\mathcal{G}$ share a common node. More formally, edge $a_{u_f,v_f} \in \{0, 1\}$, for $u_f$, $v_f \in \mathcal{V}_f$, can be expressed as

\begin{equation}
\label{eq:edge_features_links}
    a_{u_f,v_f} = \begin{cases}
            1 & \text{if $\mathbf{o}(u_f) = \mathbf{o}(v_f)$} \\
            0 & \text{otherwise}
        \end{cases}~,
\end{equation}
where $\mathbf{o}(u_f)$ denotes the node $o \in \mathcal{V}$ from which the edge $e \in \mathcal{E}$ associated with $u_f$ originates. This definition ensures that the auxiliary graph $\mathcal{G}_f$ captures the relationships between edge features that are associated with the same node in the original network.

Finally, the auxiliary GNN is tasked to determine the edges weights through the auxiliary graph as described above. As previously mentioned, this procedure involves an end-to-end learning approach. This accounts for learning the optimal edge weights (i.e., node embeddings on the auxiliary graph), given the edge features of $\mathcal{G}$, and the optimal policy parameters in a joint fashion. To meet this objective, a multi-layered GNN-based architecture is designed. First, $\mathcal{G}_f$ is processed through the auxiliary GNN to compute the edge weights for $\mathcal{G}$. Subsequently, the policy-GNN operates on $\mathcal{G}$ using the edge weights estimated by the auxiliary GNN in the previous step. Upon performing a backpropagation step, this architecture allows the concurrent update of the auxiliary GNN and the policy-GNN parameters.

% remove page numbers
\thispagestyle{empty}

\subsection{UEs' requirements and traffic modeling}
\label{sec:MARL_system_model_traffic}
In the considered system model, the user generation process follows a \ac{PCP}. A PCP is a stochastic point process defined as the union of points resulting from $M$ independent homogeneous Poisson point processes (PPPs) centered around the base stations on a Euclidean space $\mathbb{R}^2$. More formally, let $\Phi_{C_i}$ be a homogeneous PPP, centered on base station $C_i$ with intensity $\lambda_{C_i} > 0$, which generates a set of random points $\mathbf{s} \in \mathbb{R}^2$, denoted by $\mathcal{C}_{\mathbf{s}, i}$. Each user belongs to one of $S$ distinct categories based on its BER requirement. Consequently, the i-th base station $C_i$ is characterized by $S$ intensity parameters, denoted by $\lambda_{C_i}^{(k)}$, $k \in \{1, \dots, S\}$, accounting for distinct user categories. The resulting PCP, denoted by $\mathcal{U}$, is defined as the union of all resulting points
\begin{equation}
\mathcal{U} = \bigcup_{\substack{i \in \{1, \dots, M\} \\ k \in \{1, \dots, S\}}} \mathcal{C}_{\mathbf{s}, i}^{(k)}.
\end{equation}
To meet the varying BER requirements of different user categories, an independent adaptive modulation and coding scheme (MCS) is employed for each category. The adaptive MCS tailors the link spectral efficiency $\eta$ on a user basis to ensure that the required average BER is achieved for every $k$-th category. As a general expression, the relationship between the signal-to-noise ratio and average bit error rate $P_b$ for an uncoded M-QAM modulation scheme can be obtained from the union bound on error probability, which yields a closed-form
expression that is a function of the distance between signal constellation points \cite{goldsmith2005wireless}
\begin{equation}\label{eq:error_bound}
P_b \simeq \frac{1}{\log_2 L} \frac{L-1}{L}~\text{erfc}\left(\sqrt{\frac{\vert h_0\vert^2}{2\cdot N}}\right)~,
\end{equation}
where $L = \sqrt{M}$ denotes the constellation order, $2\vert h_0 \vert$ denotes the minimum distance between signal constellation points, and $N$ denotes the noise power.

\subsection{Partially Observable Markov Decision Process}

The \ac{MARL} problem is formulated as a \acf{POMDP}, where agents collect local observations from the global environment. Considering the nature of the optimization problem at hand, which involves distributed agents collectively aiming for an optimal power-tuning configuration, and because there are no temporal dependencies between the actions of the agents, the problem is formulated as a stateless \ac{POMDP}. In this formulation, the state transition dynamics are independent of past states or actions. As a consequence, the associated \ac{POMDP} tuple is given by 

\begin{equation}
    \langle \mathcal{S}, \mathcal{A}, R(s, a) \rangle~,
\end{equation}
where $\mathcal{S}$ denotes the state/observation space, $\mathcal{A}$ denotes the action space, and $R(s, a)$ indicates the reward function as a function of the state $s$ and action $a$.

\subsubsection{Observation space}
Each agent collects information pertaining to user traffic distribution across a fixed-size grid $G$ defined in polar coordinates and divided into bins, which are indexed henceforth with distance $d$ and angle $\phi$ with respect to the position of the base station. This approach results in a state space of constant dimensions that can be represented as a 3D tensor:
\begin{equation}
    \mathcal{S} = \begin{bmatrix}
        \mathbf{u}_{d_1,\phi_1} & \dots & \mathbf{u}_{d_1,\phi_n} \\
        \mathbf{u}_{d_2,\phi_1} & \ddots & \mathbf{u}_{d_2,\phi_n} \\
        \mathbf{u}_{d_m,\phi_1} & \dots & \mathbf{u}_{d_m,\phi_n} \\
    \end{bmatrix}~,
\end{equation}
where $\mathbf{u}_{d,\phi} = (u_{d,\phi}^{(1)}, \dots, u_{d,\phi}^{(S)})$ is a vector denoting the aggregated traffic for all the categories, and each $u_{d,\phi}^{(k)}$ is evaluated as

\begin{equation}\label{eq:user_counting}
    u_{d,\phi}^{(k)} = \sum_{l \in G_{d,\phi}} t_l^{(k)}~,
\end{equation}
where $t_l^{(k)}$ denotes the traffic demand of the $l$-th user of category $k$ in the bin in $G$ indexed by $(d,\phi)$.

\subsubsection{Action space}
Each agent $i$ is tailored to tune its own transmitting power $\mathbf{p}_i$, modeled as a discrete set, as a function of its own local observations. The action space is thus given by
\begin{equation}
    \mathcal{A} = \{ p_{0}, \dots, p_{M}\}~,
\end{equation}
where $M$ denotes the number of available power levels.

\subsubsection{Optimization problem}
Here, the objectives that steer the efforts of distributed agents in achieving an optimal solution are delineated. The goal is to solve the following optimization problem
\begin{equation}\label{eq:MARL_opt_problem}
    \arg \max_{\mathbf{p}} \sum_{l \in G} \sum_k \eta_{l}^{(k)}(\mathbf{p}) B_{l}(\mathbf{p})~,
\end{equation}
where $\eta_{l}^{(k)}(\mathbf{p})$ denotes the link spectrum efficiency of the $l$-th user, for all users in the reference area given by $G$. The link spectrum efficiency is evaluated as a function of the perceived \ac{SINR} and service category $k$. Furthermore, the available bandwidth for the $l$-th user, denoted by $B_{l}(\mathbf{p})$, depends on the specific scheduling mechanism in use and the total number of users connected to the base station serving the $l$-th user.

The objective function in \eqref{eq:MARL_opt_problem} encompasses two critical considerations: by choosing a proper collective power configuration $\mathbf{p}$, the base stations should maximize the average link spectral efficiency, while at the same time performing \acf{MLB} to equally balance the number of users among different base stations. User assignments to base stations occur upon executing action $a$ and following a best-server criterion. In particular, each user is assigned to the base station from which it receives the highest power. It is noteworthy that the optimization of the link spectral efficiency hinges on the distribution of user categories among base stations, which are subject to varying \ac{BER} requirements, giving rise to intricate mutual interference dynamics.

% remove page numbers
\thispagestyle{empty}

\section{Graph Multi-Agent Reinforcement Learning}
\label{sec:algo}
In the considered scenario, \ac{CTDE} framework is employed along with parameter sharing. Here, policy optimization hinges upon the application of policy gradient methods. These methods seek to maximize some scalar performance measure $J(\bm{\theta})$ of the policy parameterization, $\bm{\theta}$, through approximate gradient ascent steps:
\begin{equation}\label{eq: policy_gradient_ascent}
    \bm{\theta_}{t+1} = \bm{\theta}_t + \alpha\cdot\widehat{\nabla J(\bm{\theta}_t)}~,
\end{equation}
for some learning rate $\alpha$. The estimation of $\nabla J(\bm{\theta})$ is carried out according to the policy gradient theorem:
\begin{equation}\label{eq:policy_gradient_theroem}
\nabla J(\bm{\theta}) \propto \sum_s \mu(s) \sum_a Q_\pi(s, a) \nabla \pi(a \mid s, \bm{\theta})~,
\end{equation}
where $\pi$ denotes the policy corresponding to parameter vector $\bm{\theta}$, $\mu$ denotes the on-policy distribution under $\pi$, and $Q_\pi(s, a)$ denotes the true value function associated to state $s$ and action $a$.
Specifically, for the numerical experiments detailed in the subsequent section, the REINFORCE algorithm \cite{williams1992simple, sutton1999policy} is employed. The latter provides an approximation of the state distribution $\mu$ and the value function $Q$ defined in Eq.~\eqref{eq:policy_gradient_theroem}, through the utilization of Monte Carlo sampling. Notably, REINFORCE accounts for a model-free, on-policy approach, in which the sum over states and actions can be naturally substituted by an averaging procedure over the target policy $\pi$
\begin{equation}\label{eq:REINFORCE}
    \nabla J(\bm{\theta}) = \mathbb{E}_\pi\left[Q_\pi(S_t, A_t) \nabla \pi(a \mid S_t, \bm{\theta})\right]~.
\end{equation}
Through algebraic derivation \cite{sutton2018reinforcement}, Eq.~\eqref{eq:REINFORCE} leads to the update rule for REINFORCE, given by
\begin{equation}\label{eq:update_REINFORCE}
    \bm{\theta} \leftarrow \bm{\theta} + \alpha~\underbrace{\nabla \ln \pi(A_t \vert S_t, \bm{\theta})~R(\tau)}_{\nabla J(\bm{\theta})}~,
\end{equation}
where $R(\tau)$ denotes the return for a trajectory $\tau$ obtained by sampling over $\pi$. In the considered case, the MDP formulation is stateless, thus each episode coincides with one action step.

While the policy is acquired through centralized training, it is essential to note that during execution, the agents only have access to local information; hence, all agents operate in a distributed manner with implicit coordination. By parameterizing the policy $\pi$ using \acp{GNN}, it becomes feasible to adopt a fully decentralized execution approach, enabling the transformation and aggregation of local features from neighboring agents through the mechanism of message passing. The type of \ac{GNN} opted for the numerical experiments is the local $k$-dimensional GNN \cite{morris2019weisfeiler}. Its relative update function can be written as
\begin{equation}\label{eq:graph_conv}
\small
\mathbf{h}_v^{(l+1)}=\sigma\left(\mathbf{W}_1^{(l)}~\mathbf{h}_v^{(l)} + \mathbf{W}_2^{(l)}~\textbf{AGG}\left(\left\{e_{u, v}^{(l)} \cdot \mathbf{h}_u^{(l)}, \ \forall u \in \mathcal{N}_v \right\}\right)\right)~,
\end{equation}
where $e_{u, v}^{(l)}$ denotes the edge weight from node $u$ to node $v$, which is evaluated according to one of the strategies presented in Sec.~\ref{sec:wnet}. As confirmed by the numerical findings presented in the subsequent section, introducing distinct relational biases by modeling edge weights according to different strategies (i.e., inducing a bias on \textit{what} to communicate (message transformation), and to \textit{whom} to communicate (message passing)) has a strong influence on the collective ability to learn how to cooperate for increased performance.

% remove page numbers
\thispagestyle{empty}

\section{Simulations and Numerical Results}
\label{sec:results}

In order to test the validity of the proposed framework, and assess the role that different graph modeling strategies can have in learning a cooperative behavior, a series of numerical experiments is conducted on a simulated network scenario. The scenario comprises 11 base stations with randomly generated traffic, as described in Sec.~\ref{sec:MARL_system_model_traffic}. We considered system bandwidth $B = 60~\text{MHz}$, carrier frequency $f_c = 3.7~\text{GHz}$, and 3GPP UMa path loss channel model \cite{3GPP_38.901}. The four graph modeling strategies discussed in Sec.~\ref{sec:wnet} have been evaluated against a baseline strategy employing REINFORCE in conjunction with a policy parameterization using a classical DNN.

Specifically, the ``relation-based edges" strategy, which has been previously introduced as a general formulation, has been evaluated here according to a mutual interference criterion considering an average-to-worst-case scenario (i.e., serving cell transmitting with average tx power, interfering cell transmitting with maximum power). Assessing the effects of mutual interference relies heavily on the distribution of users across various service categories. Base stations, primarily serving users with more demanding requirements, are particularly susceptible to inter-cell interference compared to those serving other user groups. In particular, in all the training and inference tests the number of service categories $S$ is set to 3.

\subsection{Training performance}
\begin{figure}[ht]
    \centering
    \includegraphics[width = 0.93\columnwidth]{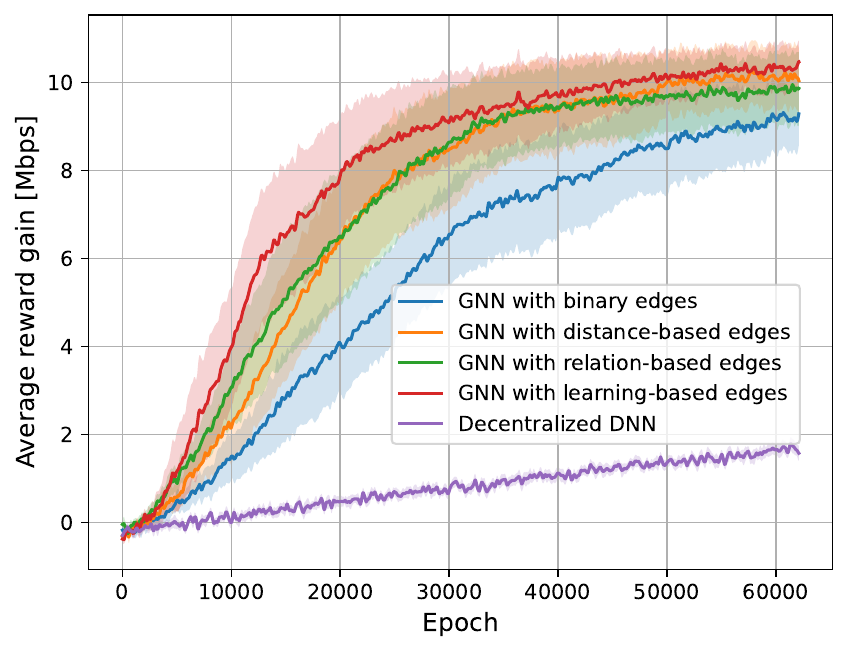}
    \caption{Training performance as a function of the number of training epochs.}
    \label{fig:training_MARL_GNN}
\end{figure}

The plot in Fig. \ref{fig:training_MARL_GNN} provides a detailed representation of the performance achieved during training time on the simulated network environment. The figure displays how the average reward evolves during training for the four considered graph models and the baseline, showcasing the performance of each model as a function of the number of training epochs. A total of 30 training instances has been carried out and results are portrayed displaying the mean reward together with the 99\% confidence intervals. As evident from the figure, employing a \ac{GNN}-based policy parameterization allows for much faster convergence and overall increased performance with respect to the solution employing a DNN-based policy. Since both methods employ CTDE with parameter sharing, the results clearly indicate that enabling agents to communicate their local features to their neighbors has a strong impact on enabling cooperation and improved performance. Also, as evident from the figure, the choice of the graph can greatly affect the performance. While the unweighted graph employing binary edges is unable to distinguish between more and less relevant neighbors, it can only achieve a mediocre performance. Vice versa, employing graph structures embedding contextual information, such as the physical proximity (orange curve) or measured level of mutual interference (green curve) into their edges, accounts for superior performance. This result shows how integrating domain knowledge in the problem formulation can drive the agents' collective behavior toward an improved solution in a shorter time. Finally, the graph with learnable edge weights (red curve) stands out as the most effective approach. By leveraging geometrical features of the environment, as described in Sec.~\ref{sec:wnet}, the proposed solution is able to learn in an end-to-end fashion the most effective formulation for edge weighting, while concurrently optimizing the weights in \eqref{eq:graph_conv} for policy parameterization.

\subsection{Generalization tests}
A unique characteristic of GNNs is their remarkable ability to generalize well to unseen scenarios, thanks to their inherent permutation equivariance (Sec.~\ref{sec:literature}). Here, results regarding the inductive capability of the proposed models are presented and discussed. As for the previous section, results are conducted over 30 independent runs. The results are presented as the mean score, accompanied by 99\% confidence intervals. These values are normalized with respect to the rewards achieved by learnable edges-based GNNs trained for the same number of epochs on larger networks.

\begin{figure}[ht]
    \centering
    \includegraphics[width = 0.93\columnwidth]{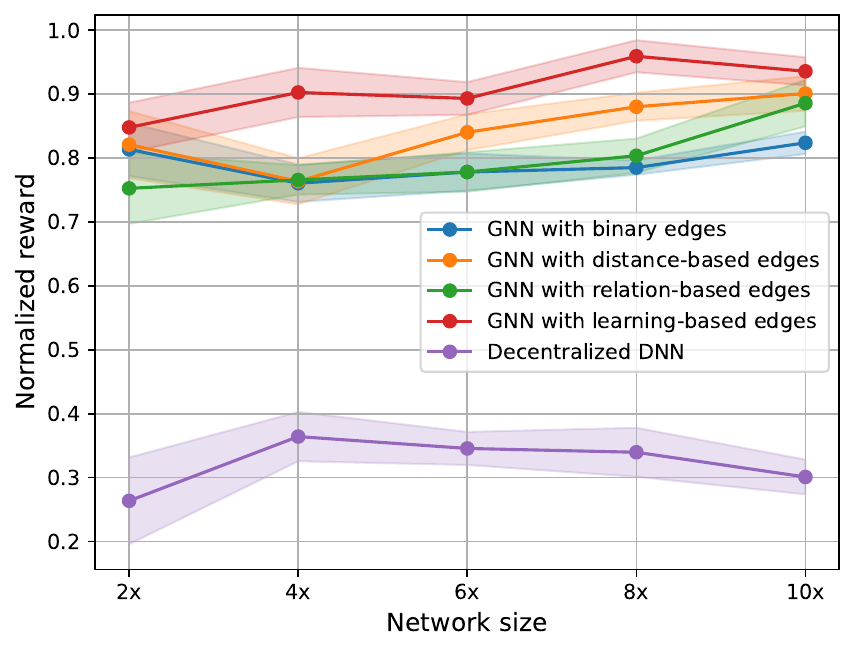}
    \caption{Generalization to networks of increasing sizes.}
    \label{fig:scabality_MARL_GNN}
\end{figure}

Fig. \ref{fig:scabality_MARL_GNN} illustrates the behavior of the learned models when they are deployed in networks of progressively larger sizes. Within this context, it becomes apparent that all GNN models exhibit a stable trend or a marginal increase in their performance as the network size increases. This behavior is attributed to the increased difficulty of the learning task as the network size grows, making it more challenging for the training to converge. Consequently, these results suggest that training on smaller scenarios effectively scales to larger, unseen ones. Remarkably, the GNN with learnable edges features better performance with respect to the other models, indicating that it not only learns a better policy but a policy that generalizes better to larger networks.

\begin{figure}[ht]
    \centering
    \includegraphics[width = 0.93\columnwidth]{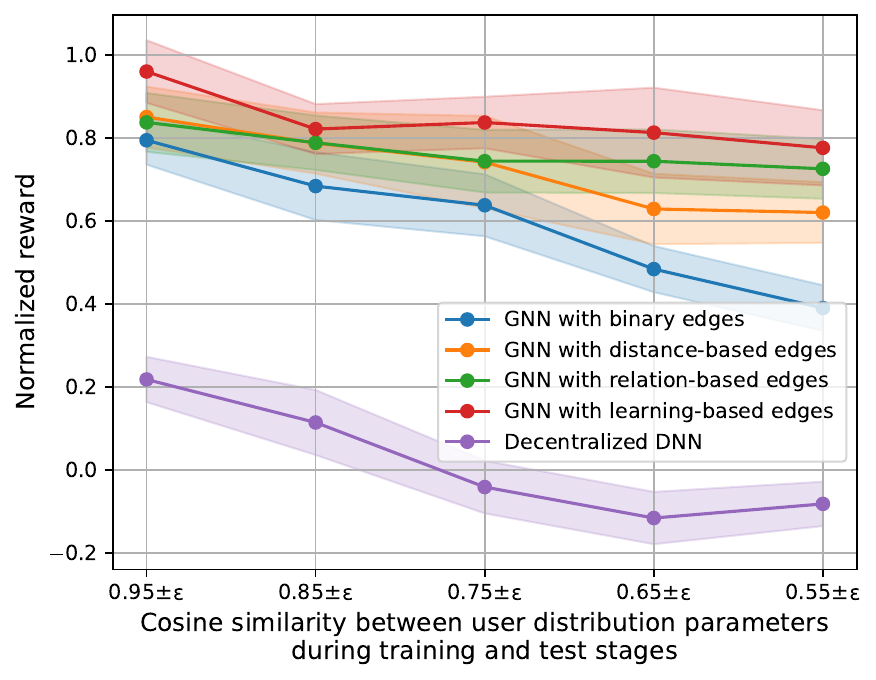}
    \caption{Generalization to networks with varying user distributions.}
    \label{fig:user_dist_MARL_GNN}
\end{figure}

As a final result, the behavior of the learned models when deployed on networks with traffic patterns that are increasingly distinct from those encountered during training is depicted in Fig. \ref{fig:user_dist_MARL_GNN}. In the simulated scenario, the traffic is modeled as a PCP, and each base station is linked with three $\lambda$ rates for each user category. To measure the difference in traffic patterns on the x-axis, the cosine similarity between the vector of $\lambda$ rates is calculated. The vectors dimension is three times the number of agents in the system. Similarly to the previous case, the results are normalized with respect to the rewards obtained by learnable edges-based GNNs trained for the same number of epochs on increasingly different traffic patterns. As shown in the figure, all GNN models demonstrate remarkable generalization capabilities, despite a marginal yet tolerable reduction in their performance as traffic patterns vary. Once more, the GNN with learnable edges exhibits superior performance.

\section{Conclusion}

In this paper, we investigated power control optimization in wireless networks through \ac{MARL} and policy parameterization with \acp{GNN}. Different adaptive graph modeling strategies are considered, including binary edges, distance-based edges, relation-based edges, and learnable edges. In particular, the latter approach, where edge weights are learned in an end-to-end fashion through the use of an auxiliary GNN, offers a promising and flexible method that can adapt to the problem at hand and simultaneously optimize the edge weights for policy parameterization. This method indeed proved to be effective by leading to faster convergence and improved performance during training. At the same time, it emerged to manage the complexity of larger and unseen inference scenarios better than baselines. Consequently, it exhibits solid generalization capabilities when addressing scalability and traffic variability challenges. To conclude, this study highlights the importance of modeling the communication structure among agents, which can significantly influence the overall performance of multi-agent systems in wireless networks.

% remove page numbers
\thispagestyle{empty}

\bibliographystyle{IEEEtran}

\bibliography{IEEEabrv,StringDefinitions,references}

\end{document}